\documentclass[pra,twocolumn,preprintnumbers,superscriptaddress]{revtex4}
\usepackage{amssymb}
\usepackage{amsmath}
\usepackage{amsfonts}
\usepackage{wasysym}
\usepackage{graphicx}
\usepackage{dcolumn}
\usepackage{bm}
\usepackage{color}
\usepackage[colorlinks,linkcolor=blue, urlcolor=blue, anchorcolor=blue, citecolor=blue]{hyperref}

\begin{document}

\title{Theory for Self-Bound States of Dipolar Bose-Einstein Condensates}
\date{\today }

\author{Yuqi Wang}
\affiliation{CAS Key Laboratory of Theoretical Physics, Institute of Theoretical Physics,
Chinese Academy of Sciences, Beijing 100190, China}
\affiliation{School of Physical Sciences, University of Chinese Academy of Sciences,
Beijing 100049, China}

\author{Longfei Guo}
\affiliation{CAS Key Laboratory of Theoretical Physics, Institute of Theoretical Physics,
Chinese Academy of Sciences, Beijing 100190, China}

\author{Su Yi}
\email{syi@itp.ac.cn}
\affiliation{CAS Key Laboratory of Theoretical Physics, Institute of Theoretical Physics,
Chinese Academy of Sciences, Beijing 100190, China}
\affiliation{School of Physical Sciences, University of Chinese Academy of Sciences,
Beijing 100049, China}
\affiliation{CAS Center for Excellence in Topological Quantum Computation, University of
Chinese Academy of Sciences, Beijing 100049, China}

\author{Tao Shi}
\email{tshi@itp.ac.cn}
\affiliation{CAS Key Laboratory of Theoretical Physics, Institute of Theoretical Physics,
Chinese Academy of Sciences, Beijing 100190, China}
\affiliation{CAS Center for Excellence in Topological Quantum Computation, University of
Chinese Academy of Sciences, Beijing 100049, China}

\begin{abstract}
We investigate the self-bound states of dipolar Dy condensates with the Gaussian-state ansatz which improves the conventional coherent-state ansatz with multimode squeezed coherent states. We show that the self-bound states consist of the experimentally observed self-bound liquid phase and the unobserved self-bound gas phase. The numerically obtained gas-liquid boundary is in good agreement with experimental data. Our theory also allows one to extract the real part of the three-body coupling constant of the Dy atoms from the particle number distribution of the condensates. In particular, we results show that the self-bound states are stabilized by the short-range three-body repulsion. Our study shed a different light to understand the self-bound droplets of Bose-Einstein condensates.
\end{abstract}

\maketitle

\textit{Introduction.}---Since the observations of the stable liquid droplets in dipolar condensates in the mean-field collapse regime~\cite{Rosensweig,singledropEr}, this novel quantum phase has received increasing attention in the community. To date, there have rapid experimental progresses in exploring the fundamental properties of these quantum droplets, such as the stabilization mechanism~\cite{manydropDy,singledropEr}, the collective excitations~\cite{singledropEr,scissorsmode}, the self-bound states~\cite{singledropDy,beyondQF}, the striped states~\cite{stripedrop}, and the metastable states with supersolid characteristics~\cite{supersolid1,supersolid2,supersolid3}. Aside from dipolar systems, liquid droplets were also observed  in attractive bosonic mixtures~\cite{2drop2d,2dropsoli-drop,2drop3d}. In addition, the droplet-droplet collisions in Bose-Bose mixtures were also experimentally studied~\cite{2dropcoll}.

From the theoretical side, the stabilization mechanism of the liquid droplets is of fundamental importance. It is now widely accepted that, unlike normal liquids, the two-body attraction is balanced at large density by quantum fluctuation~\cite{manydropDy,singledropEr}. To the first order, the quantum fluctuation can be incorporated into the Gross-Pitavskii equation (GPE) through the Lee-Huang-Yang (LHY) correction to the condensate energy~\cite{LHY} and leads to the extended Gross-Pitaevskii equation (EGPE)~\cite{dipolarTh1,dipolarTh2,twocomponentTh}. Although the EGPE explains many experimental observations, the discrepancy between the theoroetical and experimental results still exists~\cite{beyondQF}. In fact, there exists an intrinsic flaw to apply the EGPE to the droplet state as the LHY correction is obtained under the assumption that the collective excitation upon the coherent ground state is stable. Moreover, the EGPE cannot explain the asymmetric particle number distribution (PND) obtained in experiments~\cite{singledropDy,beyondQF}.

In this work, we revisit the stabilization-mechanism problem of the self-bound dipolar droplets by using the Gaussian-state theory (GST)~\cite{Shi2018,Tommaso,Shi2019} in which the many-body ground state wave functions of the condensates take the form of a multimode squeezed coherent state. As a result, the quantum fluctuation is treated in a self-consistent manner within the mean-field framework. A critical ingredient in our study is the conservative three-body interaction. The inclusion of the squeezing components allows us to obtain the unknown real part of the three-body coupling strength by fitting the experimentally measured PND~\cite{singledropDy}, which is subsequently used to map out the phase diagram on the parameter plane consisting of the relative dipolar interaction strength and the particle number. Originating from the competition between the long-range two-body attraction and the short-range three-body repulsion, there exist two self-bound phases, i.e., the self-bound liquid (SBL) and the self-bound gas (SBG) phases. Remarkably, the numerical results for the critical particle number of the SBL phase are in good agreement with the experimental values. Although the three-body repulsion was previously considered the force that balances the two-body attraction at higher densities in Refs.~\cite{Blakie2015,Xi2016,Blakie2016}, these GPE-based studies implicitly assume that the many-body ground state wave function of the condensate is a coherent state. Consequently, the estimated real part of the three-body coupling constant is much larger than the expected value~\cite{manydropDy}. While in our work, the appearance of the squeezed component in the self-bound states gives rise to unusual enhancement to the three-body coupling constant such that the two-body attraction is balanced at a rather smaller three-body coupling strength.

\textit{Formulation.}---We consider a ultracold gas of $N$ $^{164}$Dy atoms in the free space. The total Hamiltonian of the system, $H=H_{0}+H_{2B}+H_{3B}$, consists of the single-, two-, and three-particle terms. In the second-quantized form, the single-particle Hamiltonian reads $H_{0}=\int d\mathbf{r}\hat \psi ^{\dagger }(\mathbf{r})h_{0}\hat \psi (\mathbf{r})$, where $\psi (\mathbf{r})$ is the field operator and $h_{0}=-\nabla ^{2}/(2M)-\mu $ with $M$ being the mass of the atoms and $\mu$ the chemical potential. The two-particle Hamiltonian takes the form
\begin{equation}
H_{2B}=\frac{1}{2}\int d\mathbf{r}d\mathbf{r}^{\prime }\hat \psi ^{\dagger }(%
\mathbf{r})\hat \psi ^{\dagger }(\mathbf{r}^{\prime })U(\mathbf{r}-\mathbf{r}%
^{\prime })\hat \psi (\mathbf{r}^{\prime })\hat \psi (\mathbf{r}).
\end{equation}%
Here, we assume that two atoms interact via the contact and the dipolar interactions, i.e.,
\begin{align}
U({\mathbf r})=\frac{4\pi \hbar^2a_{s}\delta({\mathbf r})}{M}+\frac{\mu_0\mu^2}{4\pi}\frac{1-3\cos ^{2}\theta _{\mathbf{r}}}{r^{3}}
\end{align}
where $a_s$ is the $s$-wave scattering length, $\mu_0$ is the vacuum permeability, $\mu=9.93\,\mu_B$ is the magnetic dipole moment of Dy atom with $\mu_B$ being the Bohr magneton, and $\theta_{\mathbf r}$ is the polar angle of ${\mathbf r}$. For convenience, we introduce the relative dipolar strength $\varepsilon_{\rm dd}=\mu_0\mu^2M/(12\pi\hbar^2a_s)$. Finally, the three-body interaction term can be expressed as
\begin{equation}
H_{3B}=\frac{g_{3}}{3!}\int d\mathbf{r}\hat \psi ^{\dagger 3}(\mathbf{r})\hat \psi^{3}(\mathbf{r}),
\end{equation}%
where $g_3$ is the three-body coupling constant. Although the real part of $g_3$ is difficult to measure experimentally, as shall be shown, it can be estimated through our theory. Without loss of generality, we assume that $g_3$ is real since we shall only focus on the ground state of the system.

To proceed, let us briefly outline the GST which assumes that the many-body wave function of the condensate takes the following variational ansatz~\cite{Shi2018,Tommaso,Shi2019}
\begin{align}
\left\vert \Psi _{\mathrm{GS}}\right\rangle =e^{\hat{\Psi}^{\dagger }({\mathbf r})\Sigma
^{z}({\mathbf r},{\mathbf r}')\Phi({\mathbf r}')}e^{i\hat{\Psi}^{\dagger }({\mathbf r})\xi({\mathbf r},{\mathbf r}') \hat{\Psi}({\mathbf r}')/2}\left\vert
0\right\rangle ,  \label{VA}
\end{align}%
where the coherent part of the condensate is described by the mean value $\Phi({\mathbf r})=\langle\hat{\Psi}({\mathbf r})\rangle =\big(\phi_c(\mathbf{r}),\phi_c^{\ast }(\mathbf{r})\big)^T$ of $\hat \Psi({\mathbf r})\equiv\big(\hat \psi (\mathbf{r}),\hat \psi ^{\dagger }(\mathbf{r})\big)^{T}$ defined in the Nambu basis, $\Sigma ^{z}({\mathbf r},{\mathbf r}')=\sigma ^{z}\otimes \delta (\mathbf{r-r}^{\prime })$ with $\sigma^z$ being the Pauli matrix, and the matrix $\xi({\mathbf r},{\mathbf r}')$ characterizes the squeezing of the condensate. It should be noted that, for shorthand notation, the products in the exponents of Eq.~\eqref{VA} should be understood as the matrix multiplication in
the Nambu space and the integration in the coordinate space~\cite{SM}. To remove the redundancy~\cite{Shi2018,Tommaso}, instead of $\xi$ we introduce the covariance matrix $\Gamma (\mathbf{r,r}^{\prime })=\left\langle \{\delta
\hat \Psi (\mathbf{r}),\delta \hat \Psi ^{\dagger }(\mathbf{r}^{\prime})\}\right\rangle$ as variational parameters, where $\{\cdot,\cdot\}$ denotes the anticommutator and $\delta \hat \Psi =\hat \Psi -\Phi$ is the fluctuation field. It can be verified that $\Gamma$ and $\xi $ are connected through $\Gamma=SS^\dagger$ with $S=e^{i\Sigma ^{z}\xi}$ satisfying $S\Sigma ^{z}S^{\dagger }=\Sigma ^{z}$. Furthermore, the matrix elements of $\Gamma$ in the $2\times 2$ Nambu representation are related to the normal and the anomalous correlation functions, $G(\mathbf{r,r}^{\prime })=\left\langle \delta \hat \psi ^{\dagger }(\mathbf{r}^{\prime
})\delta \hat \psi (\mathbf{r})\right\rangle $ and $F(\mathbf{r,r}^{\prime
})=\left\langle \delta \hat \psi (\mathbf{r})\delta \hat \psi (\mathbf{r}^{\prime
})\right\rangle$, through the relations $\Gamma _{11}(\mathbf{r,r}^{\prime })=\Gamma _{22}^{T}(\mathbf{r,r}^{\prime })=2G(\mathbf{r,r}^{\prime})+\delta (\mathbf{r-r}^{\prime })$ and $\Gamma _{12}(\mathbf{r,r}^{\prime })=\Gamma _{21}^{\dagger}(\mathbf{r,r}^{\prime })=2F(\mathbf{r,r}^{\prime })$.

The Wick's theorem gives rise to the mean-field Hamiltonian~\cite{SM}
\begin{equation}
H_{\mathrm{MF}}=E_{0}+(\delta\hat \psi ^{\dagger }\eta +\mathrm{H.c.})+\frac{1}{2%
}:\delta\hat \Psi ^{\dagger }\mathcal{H}\delta\hat \Psi :  \label{Hmf}
\end{equation}
up to the quadratic orders in $\delta\hat \Psi$, where $E_{0}=\left\langle \Psi _{\mathrm{GS}}\right\vert H\left\vert \Psi_{\mathrm{GS}}\right\rangle $ is the average
energy, the symbol $::$ denotes the normal-ordering with respect to the Gaussian
state (\ref{VA}), and the derivation of $\eta[\phi,\Gamma]$ and $\mathcal{H}[\phi,\Gamma]\equiv\begin{pmatrix}
\mathcal{E}[\phi,\Gamma]&\Delta[\phi,\Gamma]\\\Delta^\dag[\phi,\Gamma]&\mathcal{E}^*[\phi,\Gamma]\end{pmatrix}$ is left in the Supplementary Materials~\cite{SM}. The ground-state solutions, $\{\phi,\Gamma\}$, can be obtained by evolving the imaginary-time equations of motion (EOM)~\cite{Shi2018,Shi2019}
\begin{subequations}
\label{Imv}
\begin{align}
\partial _{\tau }\Phi & =-\Gamma
\begin{pmatrix}
\eta \\
\eta ^{\ast }%
\end{pmatrix}%
,  \label{Imva} \\
\partial _{\tau }\Gamma & =\Sigma ^{z}\mathcal{H}\Sigma ^{z}-\Gamma \mathcal{%
H}\Gamma ,  \label{Imvb}
\end{align}
\end{subequations}
which converge for a sufficiently large imaginary time $\tau$. The numerical procedures for evolving Eq.~\eqref{Imv} are detailed in the Supplementary Materials.

\textit{General properties of the ground states.}---Before presenting the results, let us first analyze the general structure of the many-body wave function for the condensate. To this aim, we diagonalize the normal and the anomalous correlation functions into $G(\mathbf{r,r}^{\prime})\approx \sum_{j=1}^{\mathcal{S}} N_{s,j}\bar \phi_{s,j}(\mathbf{r})\bar \phi_{s,j}^{\ast }(\mathbf{r}^{\prime })$ and $F(\mathbf{r,r}^{\prime })\approx \sum_{j=1}^{\mathcal{S}} \sqrt{N_{s,j}(N_{s,j}+1)}\bar \phi_{s,j}(\mathbf{r})\bar \phi_{s,j}(\mathbf{r}^{\prime })$ (via the Takagi diagonalization), where $\mathcal{S}$ is the total number of squeezed modes, $N_{s,j}$ is the atom number in the $j$th squeezed mode $\bar \phi_{s,j}$ satisfying the orthonormal condition $\int d{\mathbf r}\bar\phi_{s,i}^*\bar\phi_{s,j}=\delta_{ij}$. Without loss of generality, we always assume that $N_{s,j}$ are sorted in descending order with respect to the index $j$. The total number of atoms in the squeezed modes is $N_s=\sum_jN_{s,j}$. Now, the ground state wave function can be expressed as
\begin{align}
\left\vert \Psi_{\mathrm{GS}}\right\rangle &=e^{\sqrt{N_{c}}(\hat b_{c}^{\dagger }-\hat b_{c})}\prod_{j=1}^{\mathcal{S}}e^{\frac{1}{2}\xi_{j}(\hat b_{s,j}^{\dagger 2}-\hat b_{s,j}^{2})}|0\rangle,\label{qust1mod}
\end{align}
where $N_c=\int d{\mathbf r}|\phi_c({\mathbf r})|^2$ is the atom number in the coherent state, $\hat b_c=\int d{\mathbf r}\hat\psi({\mathbf r})\bar\phi_c^*({\mathbf r})$ with $\bar\phi_c({\mathbf r})=\phi_c({\mathbf r})/\sqrt{N_c}$ being the normalized mode function for the coherent component, and $\xi_j=\mathrm{arcsinh}\sqrt{N_{s,j}}$ is the squeezing parameter in the mode $\hat b_{s,j}=\int d\mathbf{r}\hat \psi(\mathbf{r})\bar\phi_{s,j}^*(\mathbf{r})$.

Of particular interest, in the presence of the squeezed component, the PND and equivalently the quantum statistics of $|\Psi_{\rm GS}\rangle$ may differ from that of a coherent state significantly. To see this, we expand the coherent mode $\hat b_c$ using the squeezed modes as
\begin{align}
\hat b_{\mathrm{c}}^{\dagger}=\sum_{j=1}^{\mathcal{S}}\alpha_j\hat b_{s,j}^{\dagger}+\alpha_\perp\hat b_{\perp}^{\dagger},
\end{align}
where $\alpha_j=\int d{\mathbf r}\bar\phi_{s,j}^*({\mathbf r})\bar\phi_c({\mathbf r})$, $\alpha_\perp^2=1-\sum_j|\alpha_{j}|^2$, and $\hat b_\perp$ represents the mode that is perpendicular to all $\hat b_{s,j}$. The variational ground state~\eqref{qust1mod} can now be expressed as
\begin{eqnarray*}
|\Psi _{\mathrm{GS}}\rangle  &=&e^{\sqrt{N_{c}}\alpha _{\perp }(\hat{b}_{\mathrm{\perp }}^{\dagger }-\hat{b}_{\mathrm{\perp }})}\prod_{j=1}^{\mathcal{S}}
e^{\sqrt{N_{c}}(\alpha _{j}\hat{b}_{s,j}^{\dagger }-\alpha _{j}^{\ast }\hat{b}_{s,j})} \\
&&\times e^{\frac{1}{2}\xi _{j}(\hat{b}_{s,j}^{\dagger 2}-\hat{b}_{s,j}^{2})}\left\vert 0\right\rangle ,
\end{eqnarray*}
The PND for mode $\hat b_\perp$ is $p_\perp(\ell)=e^{-N_c |\alpha_\perp|^2}(N_c|\alpha_\perp|^2)^{\ell}/\ell!$ and that for mode $\hat b_{s,j}$ is~\cite{quoptics}
\begin{align}
p_{s,j}(\ell)&=\frac{(\frac{1}{2}\tanh\xi_j)^\ell}{\ell!\cosh\xi_j}\left|H_\ell\left[\gamma_j(\sinh 2\xi_j)^{-1/2}\right]\right|^2\nonumber\\
&\quad\times\exp\left[-|\alpha_j|^2-\frac{1}{2}(\alpha_j^{*2}+\alpha_j^2)\tanh\xi_j\right],
\end{align}
where $\gamma_j=\alpha_j\cosh\xi_j+\alpha_j^*\sinh\xi_j$ and $H_\ell(x)$ are Hermite polynomials. The PND of $|\Psi_{\rm GS}\rangle$ can then be expressed into a recursive form as follows. Let $P_j(n)$ denote the PND of the state containing first $j$ squeezed modes, the PND containing first $j+1$ squeezed modes can then be expressed as
\begin{align}
P_{j+1}(n)=\sum_{\ell=0}^nP_j(n-\ell)p_{s,j+1}(\ell),
\label{rr}
\end{align}
where $P_0(n)\equiv p_\perp(n)$. Applying the relation (\ref{rr}) successively will eventually leads to the total PND $P_\mathcal{S}(n)$.

In general, $P_\mathcal{S}(n)$ is peaked at roughly $n\approx N_c$. To the left of this peak, the PND is mainly determined by Poission distribution of the coherent state such that $P_\mathcal{S}(n)$ increase abruptly as $n$ approaches $N_c$. While to the right of the peak, $P_\mathcal{S}(n)$ has a long tail at large $n$~\cite{quoptics}. As a result, $P_\mathcal{S}(n)$ becomes asymmetric around $N_c$, which is in striking contrast to the PND of a coherent state.

\begin{figure}[ptb]
\centering
\includegraphics[width=0.8\columnwidth]{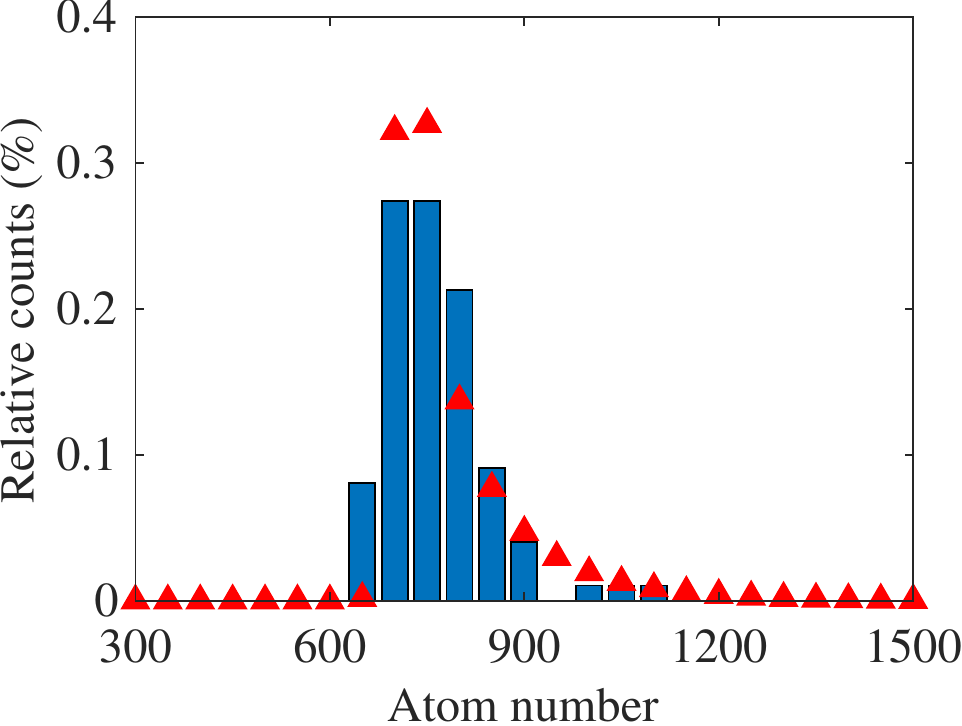}
\caption{(color online). Read triangles shows the PND for the state obtained with the parameters $N^*=790$ and $g_3^*=6.73\times 10^{-40}\hbar {\mathrm m}^6/\mathrm{s}$ at magnetic field $B=6.624\,{\mathrm G}$. The histogram represent the experimental data~\cite{singledropDy}.}\label{fitpnd}
\end{figure}

\textit{Determination of $g_3$.}---In Ref.~\cite{singledropDy}, the PNDs of the condensates were experimentally measured under various magnetic fields, or equivalently, $\varepsilon_{\rm dd}$'s. Because the measured PNDs were asymmetric around the mean particle number, they were fitted with a function that is the convolution of a Gaussian and a Maxwell-Boltzmann distribution in Ref.~\cite{singledropDy}. Here we attribute the measured PNDs to that of the Gaussian state $|\Psi_{\rm GS}\rangle$ and fit the measured PNDs with $P_\mathcal{S}(n)$. Specifically, for every PND obtained under a given magnetic field in Ref.~\cite{singledropDy}, we have carried out extensive fittings by systematically varying the values of $N$ and $g_3$. It is found that the experimentally measured PNDs are broader than that obtained with $P_\mathcal{S}(n)$, which might originate from the detection noise and the thermal fluctuations of the condensates in the evaporation. The best fit, as shown in Fig.~\ref{fitpnd}, is achieved under the magnetic field $B=6.624\,{\rm G}$ with fitting parameters $N^*=790$ and $g_3^*=6.73\times 10^{-40}\hbar {\mathrm m}^6/\mathrm{s}$, where $N^*$ roughly agrees with the experimentally measured mean number of atoms. Interestingly, the value of $g_3^*$ is about one order of magnitude smaller that used in previous work~\cite{Blakie2015} and is therefore more favorable~\cite{manydropDy}. Unless otherwise stated, the value of the three-body coupling strength will always be fixed to $g_3^*$ in below.

\begin{figure}[ptb]
\centering
\includegraphics[width=0.99\columnwidth]{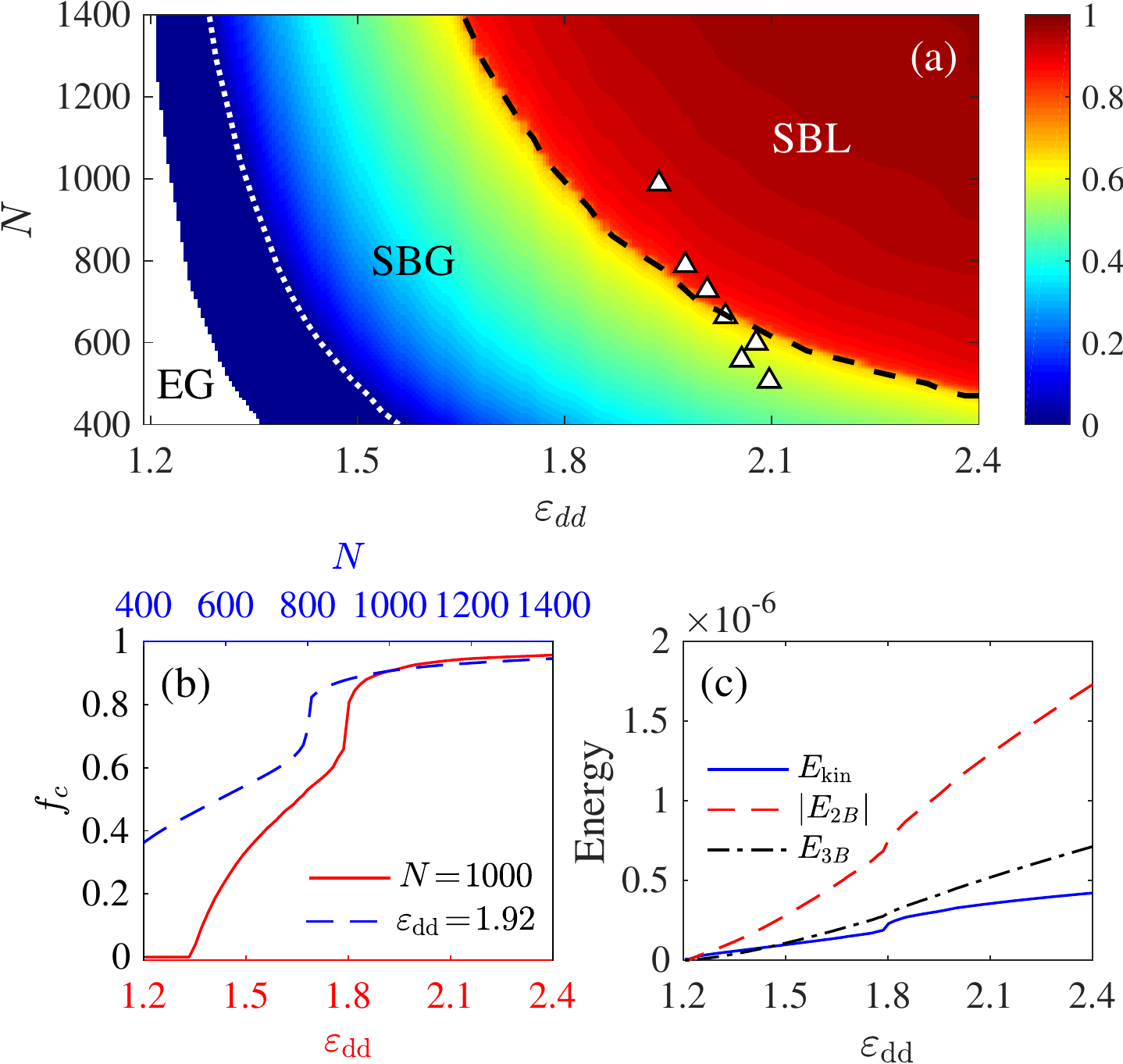}
\caption{(color online). (a) Distribution of the coherent-component fraction on the $\varepsilon_{\rm dd}N$ parameter plane. The black dashed line marks the boundary across which the value of $f_c$ changes abruptly. The white dash-dotted line denote boundary between the self-bound state and the expanding state. Empty triangles represent the experimental data for the critical particle number. (b) The solid line shows the $\varepsilon_{\rm dd}$ dependence of $f_c$ for $N=1000$ and dashed line plots the $N$ dependence of $f_c$ for $\varepsilon_{\rm dd}=1.92$. (c) The $\varepsilon_{\rm dd}$ dependence of $E_{\rm kin}$, $|E_{2B}|$, and $E_{3B}$ for $N=1000$.}\label{phase}
\end{figure}

\begin{figure}[ptb]
\centering
\includegraphics[width=0.95\columnwidth]{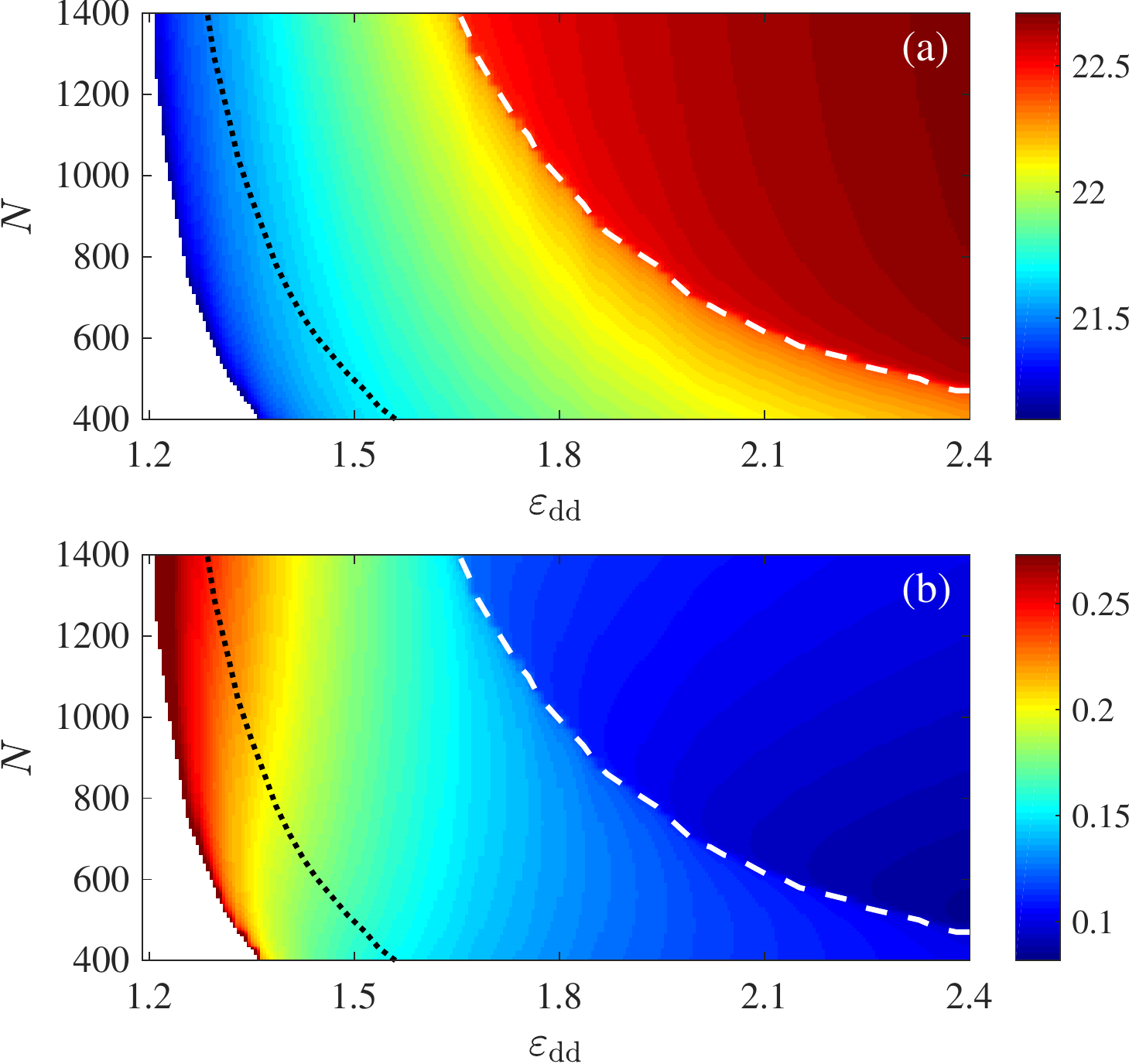}
\caption{(color online). Distribution of the peak density $\log n_{\rm peak}$ and the radial width $\sigma_\rho$ of the condensate on the $\varepsilon_{\rm dd}N$ plane. Here $n_{\rm peak}$ is in units of ${\rm m}^{-3}$ and $\sigma_\rho$ is in units of $\mu{\rm m}$}\label{denwid}
\end{figure}

\textit{Phase diagram.}---In Fig.~\ref{phase}(a), we map out the coherent-component fraction $f_c\equiv N_c/N$ of condensate in the $\varepsilon_{\rm dd}N$ parameter plane. To reveal more details, we plot the typical $\varepsilon_{dd}$ and $N$ dependence of $f_c$ in Fig.~\ref{phase}(b) and the $\varepsilon_{\rm dd}$ dependence of the kinetic energy $E_{\rm kin}$, the two-body interaction energy $E_{2B}$, and the three-body interaction energy $E_{3B}$ in Fig.~\ref{phase}(c). The detailed expressions for these energy components can be found in the Supplementary Materials~\cite{SM}. We note that for the parameters covered by our numerical calculations, we always have $E_{2B}<0$. Moreover, since the peak density $n_{\rm peak}$ and the radial $1/e^2$ width $\sigma_\rho$ (or the axial $1/e^2$ width $\sigma_z$) are of interest to experimentalists, we also map out the values of $\log n_{\rm peak}$ and $\sigma_\rho$ on the $\varepsilon_{\rm dd}N$ parameter plane in Fig.~\ref{denwid}. An immediate observation is that all three figures share the same structure. Namely, the parameter plan is divided into three regions representing the expanding-gas (EG) phase, the SBG phase, and the SBL phase. The condensate of the EG phase is not a self-bound state, it is therefore not of interest to the present work.

\begin{figure}[ptb]
\centering
\includegraphics[width=0.95\columnwidth]{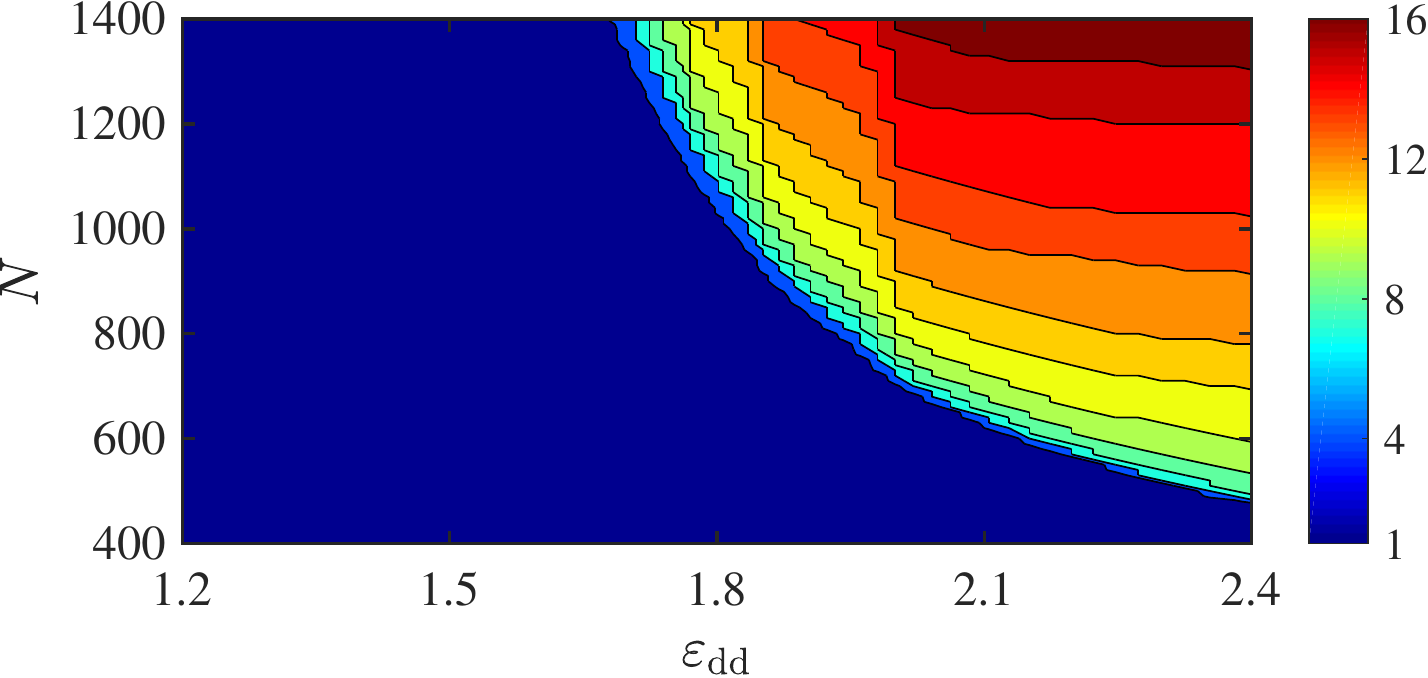}
\caption{(color online). Contour plot for the number of the NPSMs.}\label{squmodes}
\end{figure}

In the SBG phase, the attractive interaction or the atom number is large enough for the condensate to form a self-bound state. Because the density is still relatively low [see Fig.~\ref{denwid}(a)], the condensate remains in the gas phase. A important feature of the SBG phase is that the squeezed component is always dominated by a single squeezed mode (see Fig.~\ref{squmodes}). More interestingly, the SBG phase can be further divided into two subregions by the dotted line in Fig.~\ref{phase}(a): i) To the left of this boundary, the coherent-component fraction vanishes completely such that the state of the condensate is a single-mode squeezed vacuum, in analog to that of a trapped weakly attractive condensate~\cite{Shi2019}. ii) To the right of the dotted line, the ground state of the condensate is a squeezed coherent state in which $f_c$ can be as high as around $0.64$. Analyzing the $\varepsilon_{\rm dd}$ dependence of the ground-state energy $E_0$ ($=E_{\rm kin}+E_{2B}+E_{3B}$) further reveals that the transition between the squeezed vacuum and the squeezed coherent states is of third order, since the third derivative of $E_0$ with respect to $\varepsilon_{\rm dd}$ is discontinuous. This transition reflects the fact that the appearance of the coherent component breaks the $\rm{Z}_2$ symmetry possessed by the pure squeezed states. Physically, increasing $\varepsilon_{\rm dd}$ or $N$ enhances both two-body and three-body interactions, which shrinks the condensate and leads to a higher density (see Fig.~\ref{denwid}). It turns out that the coherent condensate is more energetically favorable in the high density regime, where the squeezed atoms are turned into the coherent ones. This mechanism can be further confirmed by deliberately lowering the value of $g_3$ in numerical calculations. It is found that for the same set of $\varepsilon_{\rm dd}$ and $N$, the coherent-component fraction monotonically increases when $g_3$ is lowered.

\begin{figure}[ptb]
\centering
\includegraphics[width=0.95\columnwidth]{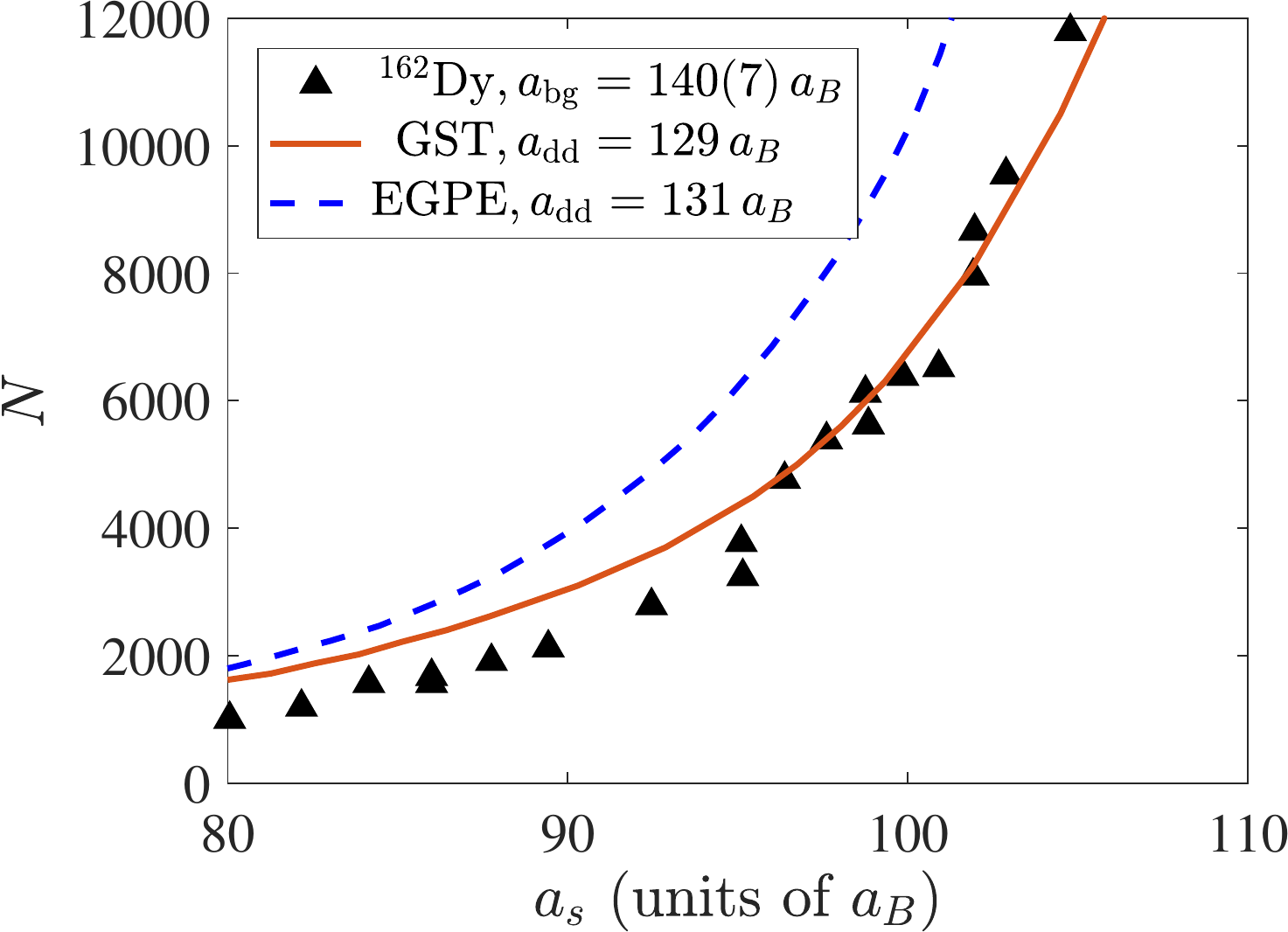}
\caption{(color online). Critical atom number for the SBL phase of the ${}^{162}$Dy condensates. Filled triangles and dashed line (both are taken from Ref.~\cite{beyondQF}) represent the experimental data and the EGPE result, respectively. Solid line is computed with the GST, in which we have used the same three-body coupling constant as that of ${}^{164}$Dy atoms.}\label{dy162}
\end{figure}

As one further increases $\varepsilon_{\rm dd}$ or $N$, the three-body repulsion becomes determinative such that $f_c$ experiences an abrupt increase and is now dominant in the condensate ($f_c\apprge0.8$). The system then enters the SBL phase through a first-order phase transition where, as shown in Fig.~\ref{denwid}, the $n_{\rm peak}$ ($\sigma_\rho$) of the condensate is significantly increased (reduced) compared that in the SBG phase. Remarkably, the SBG-SBL phase boundary is in good agreement with the experimentally measured critical atom number~\cite{singledropDy}. As shown in Fig.~\ref{dy162}, we have also numerically computed the critical atom number for the recent $^{162}$Dy experiment~\cite{beyondQF}, which is in very good agreement with experimental results compared to EGPE with LHY corrections.

To gain more insight into the gas-liquid transition, we examine the number of the squeezed modes by introducing the so-called notably-populated squeezed modes (NPSMs). Specifically, the $j$th squeezed mode is an NPSM if $N_{s,j}/N_s\geq 0.5\%$. In Fig.~\ref{squmodes}, we present the distribution of the number of the NPSMs in the $\varepsilon_{\rm dd}N$ plane, where the SBL-SBG boundary is sharply defined. The SBG phase is dominated by a single squeezed mode; while across the gas-liquid boundary, the coherent component becomes dominant and the number of NPSMs increases abruptly. Physically, the squeezing in the SBG phase originates from the overall attractive two-body interactions, same as that in the condensate with pure $s$-wave interaction~\cite{Shi2019}. On the other hand, the squeezing in the SBL phase represents the depletion of the condensate that is induced by the three-body repulsion. Indeed, it is verified that the fraction of the squeezed component increases if we increase $g_3$ or, equivalently, decrease $\varepsilon_{\rm dd}$.

\begin{figure}[ptb]
\centering
\includegraphics[width=0.95\columnwidth]{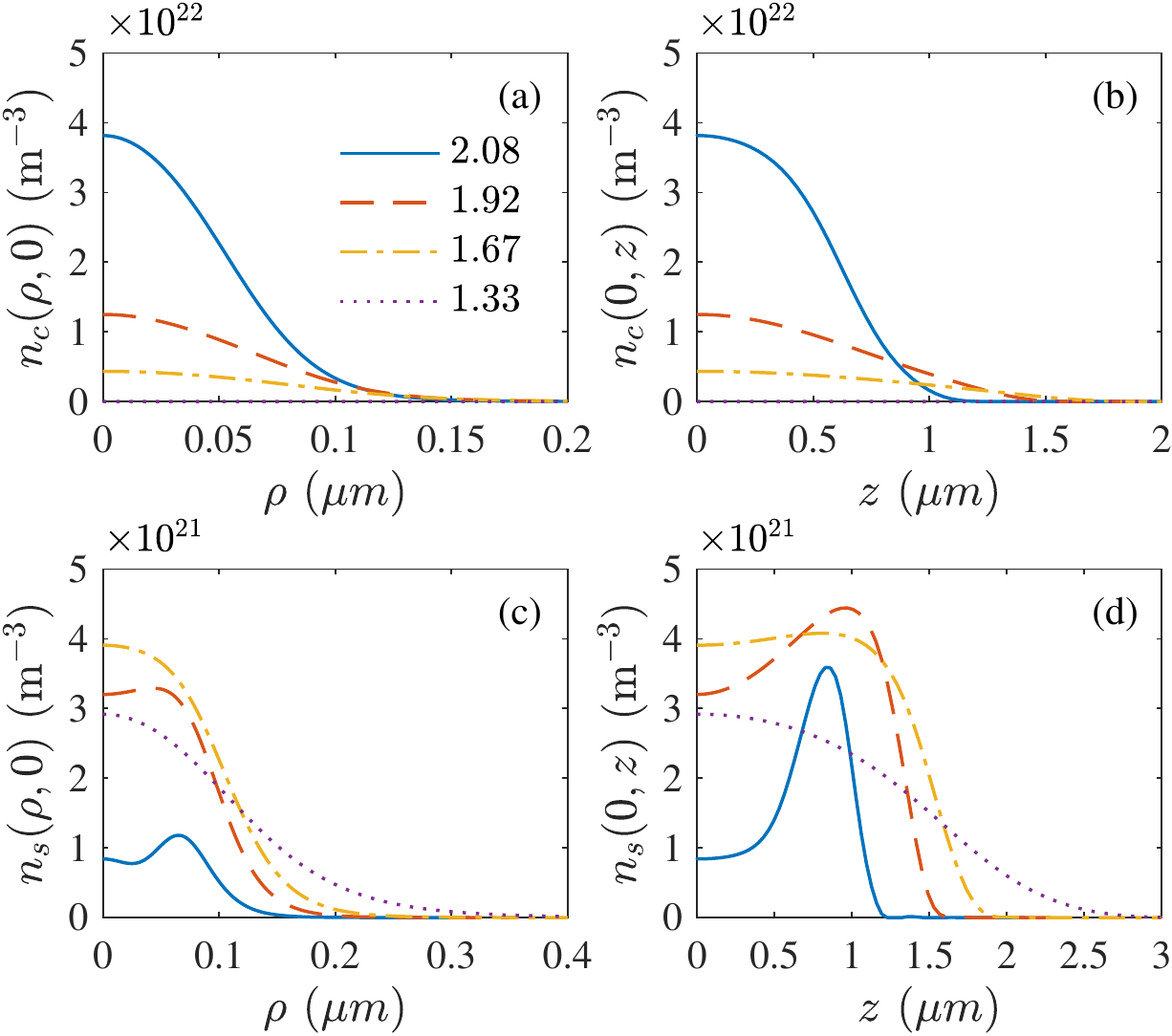}
\caption{(color online). Density profiles $n_c(\rho,0)$ (a), $n_c(0,z)$ (b), $n_s(\rho,0)$ (c), and $n_s(0,z)$ (d) for $N=700$ and $\varepsilon_{\rm dd}=2.08$ (solid lines), $1.92$ (dashed lines), $1.67$ (dash-dotted lines), and $1.33$ (dotted lines).}\label{wave}
\end{figure}

\textit{Density profiles.}---We now examine the density profiles which consists of the coherent part $n_c({\mathbf r})=|\phi_c({\mathbf r})|^2$ and the squeezed part $n_s({\mathbf r})=G({\mathbf r},{\mathbf r})$. Due to axial symmetry of the condensate, both $n_c$ and $n_s$ are cylindrically symmetric. In Fig.~\ref{wave}, we plot the density profiles along the radial $\rho=\sqrt{x^2+y^2}$ and the $z$ directions for both coherent and squeezed components with $N=700$ and for various $\varepsilon_{\rm dd}$'s. In addition, because the density is symmetric about the $z=0$ plane, only the density profiles for $z\geq 0$ are plotted.

For the coherent component, $n_c$ always reaches its peak value at the center of the condensate. It then monotonically decreases along both the radial and the axial directions. In addition, as $\varepsilon_{\rm dd}$ increases, the peak value of $n_c$ also grows along with the increase of $f_c$. As to the squeezed component, similar to the behavior of $n_c$, $n_s$ also reaches its peak at the condensate center when $\varepsilon_{\rm dd}$ is small. However, for large $\varepsilon_{\rm dd}$, the value of $n_s$ at the condensate center drops such that both $n_s(\rho,0)$ and $n_s(0,z)$ are not monotonic. Physically, this can be understood by noting the total density, $n_c+n_s$, is always maximized at the center of the condensate. As a result, the three-body repulsion is strongest at the center of the condensate which locally turns the squeezed component into the coherent one. Now, the structure of the a liquid droplet can be roughly described as follows: the high-density coherent component forms the liquid core of the condensate; while the low-density squeezed component constitutes a gas shell surrounding the liquid core.

\textit{Effective equations for the self-boudn states.}---In principle, we may derive effective equations for the ground-state spatial modes $\phi_c$ and $\{\phi_{s,j}\}_{j=1}^{\mathcal{S}}$ by minimizing the ground-state energy $E_0$ under the constraints that the unnormalized squeezed modes $\phi_{s,j}=\sqrt{N_{s,j}}\bar\phi_{s,j}$ are mutually orthogonal. This however becomes very cumbersome deep inside the SBL phase since the number of the squeezed modes could be very large. We notice that deeply inside the SBL phase most of atoms are in the coherent state, while in the SBG phase the squeezed component is dominated by a single mode, i.e., $\mathcal{S}=1$. As a result, in the deep SBL phase, the ground state is described by a single macroscopic wavefunction $\phi_c({\mathbf r})$  that obeys the conventional GPE. In the SBG phase, both coherent and squeezed condensates contain the macroscopic number of particles, which are characterized by two macroscopic wavefunctions $\chi({\mathbf r})\equiv(\phi_c({\mathbf r}),\phi_s({\mathbf r}))^T$, where $\phi_s({\mathbf r})\equiv\phi_{s,1}({\mathbf r})$ describes the mode with largest squeezing. The conditions $\delta E_0/\delta\chi^\dagger=0$ then give rise to the squeezed-coherent-state GPEs
\begin{align}
\left[h_0I+{\mathcal M}^{(2)}({\mathbf r})+\frac{g_3}{2}{\mathcal M}^{(3)}({\mathbf r})\right]\chi({\mathbf r})=0,\label{gpelike}
\end{align}
where $I$ is a $2\times 2$ identity matrix,
\begin{align}
{\mathcal M}^{(2)}({\mathbf r})&=\int d{\mathbf r}'U({\mathbf r}-{\mathbf r}')\begin{pmatrix}n_{1}({\mathbf r}') & n_{\mathrm{sc}}({\mathbf r}') \\n_{\mathrm{sc}}({\mathbf r}') & n_{3}({\mathbf r}')\end{pmatrix}
\end{align}
and
\begin{align}
{\mathcal M}^{(3)}({\mathbf r})&=\begin{pmatrix}n_{3}^{2}+\phi_{s}^{\ast 2}\phi_{c}^{2}&3n_{2}\phi_{s}\phi_{c}^{\ast } \\
2n_{9/2}\phi_{c}\phi_{s}^{\ast}&3n_{1}n_{5}+3\phi_{c}^{\ast 2}\phi_{s}^{2}
\end{pmatrix}
\end{align}
are the mean fields induced by the two- and three-body interactions, respectively. Here we have introduced the notations $n_{q}=|\phi_c|^2+q|\phi_s|^2$ and $n_{\mathrm{sc}}=2{\rm Re}(\phi_{s}^{\ast }\phi_{c})$. We note that the diagonal elements of  ${\mathcal M}^{(2)}$ are contributed by the intra-state interaction and by inter-state Hartree interaction; while the off-diagonal terms originate from the inter-state Fock and pairing interactions.

To appreciate the physical significance of Eqs.~\eqref{gpelike}, let us consider the simplest cases by assuming that the condensate is either in a pure coherent state or in a single-mode squeezed vacuum state. For either case, Eqs.~\eqref{gpelike} reduce to a unified form:
$\big[h_0+c_2\int d{\mathbf r}'U({\mathbf r}-{\mathbf r}')|\varphi({\mathbf r}')|+c_3\frac{g_3}{2}|\varphi({\mathbf r})|^4\big]\varphi({\mathbf r})=0$, where $c_2=c_3=1$ if $\varphi\equiv\phi_c$ and $c_2=3$ and $c_3=15$ if $\varphi\equiv\phi_s$. As can be seen, although both two- and three-body interaction strengths of the squeezed state are enhanced compared to those of the coherent state, the three-body interaction strength gains more increment. As shown in the expression of $E_{3B}$ in the Supplemental Material~\cite{SM}, this unusually large enhancement to the three-body interaction strength originates from the fact that the combination of squeezing and three-body interaction yields more terms when applying the Wick's theorem.

We have performed extensive numerical calculations to check the validity of these effective equations. It is found that away from the gas-liquid boundary the results obtained with the effective equations are in perfect agreement with those computed using Eqs.~\eqref{Imv}. While close to the gas-liquid boundary, because number of the squeezed modes is rapidly increased, the effective equations becomes less accurate.

\textit{Discussion and conclusion.}---As analyzed in our previous work~\cite{Shi2019}, if we sequentially solve Eqs.~\eqref{Imva} and \eqref{Imvb} iteratively starting with a pure coherent state, the EGPE is merely the resulting equation after the first iteration. Therefore, EGPE is applicable only when the condensate is dominated by the coherent component, which excludes its application to the SBG phase. Even for the coherent-component-dominant SBL phase one should be very careful about the validity of EGPE. In fact, at its current form, the EGPE only takes into account the quantum fluctuation induced by the two-body interactions; while the quantum fluctuation in the SBL phase is induced by the three-body repulsion. To further confirm this, we switch off the three-body repulsion by setting $g_3=0$ in numerical calculations, it is found that condensates always collapse for all parameters covered by Fig.~\ref{phase}(a). This result clearly suggests that the quantum fluctuation associated with two-body interactions is insufficient to balance the two-body attractions.

We note that three-body repulsion was previously considered as the source that stabilizes the two-body attraction in Refs.~\cite{Blakie2015,Xi2016,Blakie2016}. However, to match the experimental measurements, the estimated three-body coupling constant using the GPE is much higher than the expected value~\cite{singledropDy}. Within the GST, the appearance of the squeezed component leads to the unusual enhancement to the three-body coupling constant such that the two-body attraction is balanced at a rather smaller $g_3$. For parameters deep inside the SBL phase where the squeezing becomes negligibly small, it has been verified that the GPE with three-body repulsion can lead to the same results as those obtained with Eqs.~\eqref{Imv}.

In summary, we have studied the self-bound states of dipolar condensates by using the GST which allows us to treat the quantum fluctuation in a self-consistent manner. Aside from the self-bound liquid droplet phase, we predict the existence of the SBG phase. We have also shown that the gas-liquid transition is induced by competition between the long-range two-body attraction and the short-range three-body repulsion. Unlike the conventional theory based on coherent states, squeezed component emerges in the ground states of both phases with distinct physical origins: it is induced by the two-body attraction (three-body repulsion) in the SBG (SBL) phase. The appearance of the squeezed component in the self-bound states is crucial as (particularly in the SBL phase) it allows the three-body repulsion to balance the two-body attraction even with a relatively small three-body coupling constant. For experimental detections, we note that the squeezed-component-dominant SBG phase can be detected by measuring the second-order correlation function of the gas as it exhibits distinct quantum statistics from that of the coherent-component-dominant SBL phase~\cite{Shi2019}. In the future, we shall further study the dynamical formation of the self-bound states. We believe our study opens up a broad avenue to explore the brand new macroscopic quantum states in cold atom systems.

\textit{Acknowledgement.}---We thank the enlightening discussion with J. Ignacio Cirac, Eugene Demler, Carlos Navarrete-Benlloch, Li You, Chang-Pu Sun, Han Pu, Peter Zoller, Frank Wilczek, Cristiane Morais Smith, Andreas Hemmerich, Zi Cai, Tilman Pfau, and Hans Peter Buechler. TS acknowledges the Thousand-Youth-Talent Program of China. This work was supported by the NSFC (Grants No. 11434011, No. 11674334, and No. 11974363), by the Strategic Priority Research Program of CAS (Grant No. XDB28000000), and by National Key Research and Development Program of China (Grant No. 2017YFA0304501).

\onecolumngrid
\clearpage

\begin{center}
\textbf{\large Supplemental Material}
\end{center}

\setcounter{equation}{0} \setcounter{figure}{0} 
\makeatletter

\renewcommand{\thefigure}{SM\arabic{figure}} \renewcommand{\thesection}{SM%
\arabic{section}} \renewcommand{\theequation}{SM\arabic{equation}}

This supplemental material is divided in three sections. In Sec. SM1, we derive
the mean-field Hamiltonian, i.e., Eq. (\ref{Hmf}) in the main text. In Sec.
SM2, the EOM in the imaginary time evolution is projected in the angular
momentum basis $e^{im\varphi }/\sqrt{2\pi }$, which can be solved
efficiently by the Hankel transformation. In Sec. SM3, we derive the coupled
GP-like equation for coherent and squeezed condensates by minimizing the
ground state energy.

\section{Mean-field Hamiltonians}

Here we outline the derivation of the mean-field Hamiltonian (\ref{Hmf}). To this end, we substitute the expansion $\psi (\mathbf{r})=\phi (\mathbf{r})+\delta \psi (\mathbf{r})$ into the total Hamiltonian $H$, which leads to
\begin{align}
H_{0}&=\int d\mathbf{r}\Big\{\phi ^{\ast }(\mathbf{r})h_{0}\phi (\mathbf{r}%
)+\left[\delta \psi ^{\dagger }(\mathbf{r})h_{0}\phi (\mathbf{r})+\mathrm{H.c.}\right]+\delta \psi ^{\dagger }(\mathbf{r})h_{0}\delta \psi (\mathbf{r})\Big\},\\
H_{2B} &=\int d\mathbf{r}d\mathbf{r}^{\prime }U(\mathbf{r}-\mathbf{r}^{\prime })\bigg\{\frac{1}{2}\left\vert \phi (\mathbf{r})\right\vert^{2}\left\vert \phi (\mathbf{r}^{\prime })\right\vert ^{2}+\left[\left\vert \phi (%
\mathbf{r}^{\prime })\right\vert ^{2}\phi (\mathbf{r})\delta \psi ^{\dagger
}(\mathbf{r})+\mathrm{H.c.}\right] \notag \\
&\quad+\left\vert \phi (\mathbf{r}^{\prime })\right\vert ^{2}\delta \psi
^{\dagger }(\mathbf{r})\delta \psi (\mathbf{r}) +\phi ^{\ast }(\mathbf{r}%
^{\prime })\phi (\mathbf{r})\delta \psi ^{\dagger }(\mathbf{r})\delta \psi (%
\mathbf{r}^{\prime })+\frac{1}{2}\left[\phi (\mathbf{r}^{\prime })\phi (\mathbf{r}%
)\delta \psi ^{\dagger }(\mathbf{r})\delta \psi ^{\dagger }(\mathbf{r}%
^{\prime })+\mathrm{H.c.}\right]  \notag \\
&\quad+\left[\phi (\mathbf{r})\delta \psi ^{\dagger }(\mathbf{r})\delta \psi
^{\dagger }(\mathbf{r}^{\prime })\delta \psi (\mathbf{r}^{\prime })+\mathrm{%
H.c.}\right]+\frac{1}{2}\delta \psi ^{\dagger }(\mathbf{r})\delta \psi ^{\dagger }(%
\mathbf{r}^{\prime })\delta \psi (\mathbf{r}^{\prime })\delta \psi (\mathbf{r%
})\bigg\},\\
H_{3B} &=g_{3}\int d\mathbf{r}\bigg\{\frac{1}{3!}\left\vert \phi (\mathbf{r}%
)\right\vert ^{6}+\frac{1}{2}\left[\left\vert \phi (\mathbf{r})\right\vert
^{4}\phi (\mathbf{r})\delta \psi ^{\dagger }(\mathbf{r})+\mathrm{H.c.}\right]\nonumber\\
&\quad+\frac{3}{2}\left\vert \phi (\mathbf{r})\right\vert ^{4}\delta \psi ^{\dagger
}(\mathbf{r})\delta \psi (\mathbf{r})+\frac{1}{2}\left[\left\vert \phi (\mathbf{r}
)\right\vert ^{2}\phi ^{2}(\mathbf{r})\delta \psi ^{\dagger 2}(\mathbf{r})+%
\mathrm{H.c.}\right]\nonumber\\
&\quad +\frac{1}{3!}\left[\phi ^{3}(\mathbf{r})\delta \psi ^{\dagger 3}(\mathbf{r}
)+9\left\vert \phi (\mathbf{r})\right\vert ^{2}\phi (\mathbf{r})\delta \psi
^{\dagger 2}(\mathbf{r})\delta \psi (\mathbf{r})+\mathrm{H.c.}\right]  \notag \\
&\quad +\frac{1}{2}\left[\phi ^{2}(\mathbf{r})\delta \psi ^{\dagger 3}(\mathbf{r}%
)\delta \psi (\mathbf{r})+3\left\vert \phi (\mathbf{r})\right\vert
^{2}\delta \psi ^{\dagger 2}(\mathbf{r})\delta \psi ^{2}(\mathbf{r})+\phi
^{\ast 2}(\mathbf{r})\delta \psi ^{\dagger }(\mathbf{r})\delta \psi ^{3}(%
\mathbf{r})\right]\nonumber\\
&\quad+\frac{1}{2}\left[\phi (\mathbf{r})\delta \psi ^{\dagger 3}(\mathbf{r})\delta \psi ^{2}(%
\mathbf{r})+\mathrm{H.c.}\right]+\frac{1}{3!}\delta \psi ^{\dagger 3}(\mathbf{r})\delta \psi
^{3}(\mathbf{r})\bigg\}.
\end{align}
After applying the Wick's theorem with respect to the Gaussian state $\left\vert \Psi _{\mathrm{GS}}\right\rangle$, we obtain the mean-field Hamiltonian
\begin{equation}
H_{\mathrm{MF}}=E_{0}+(\delta \psi ^{\dagger }\eta +\mathrm{H.c.})+\frac{1}{2%
}\text{:}\delta \Psi ^{\dagger }\mathcal{H}\delta \Psi \text{:},
\end{equation}
where $E_0=E_{\rm kin}+E_{2B}+E_{3B}$ is the average energy of the system with
\begin{align}
E_{\rm kin}&=\int d\mathbf{r}\left[\phi ^{\ast }(\mathbf{r})h_{0}\phi (\mathbf{r}%
)+\left\langle \delta \psi ^{\dagger }(\mathbf{r})h_{0}\delta \psi (\mathbf{r%
})\right\rangle\right],\nonumber\\
E_{2B} &=\int d\mathbf{r}d\mathbf{r}^{\prime }U(\mathbf{r}-\mathbf{r}%
^{\prime })\bigg\{\frac{1}{2}\left\vert \phi (\mathbf{r})\right\vert
^{2}\left\vert \phi (\mathbf{r}^{\prime })\right\vert ^{2}+\left\langle
\delta \psi ^{\dagger }(\mathbf{r})\delta \psi (\mathbf{r})\right\rangle
\left[\left\vert \phi (\mathbf{r}^{\prime })\right\vert ^{2}+\frac{1}{2}%
\left\langle \delta \psi ^{\dagger }(\mathbf{r}^{\prime })\delta \psi (%
\mathbf{r}^{\prime })\right\rangle \right] \\
&\quad+\left\langle \delta \psi ^{\dagger }(\mathbf{r})\delta \psi (\mathbf{r}%
^{\prime })\right\rangle \left[\phi ^{\ast }(\mathbf{r}^{\prime })\phi (\mathbf{r}%
)+\frac{1}{2}\left\langle \delta \psi ^{\dagger }(\mathbf{r}^{\prime
})\delta \psi (\mathbf{r})\right\rangle \right]  \notag \\
&\quad+\text{Re}\left\langle \delta \psi
^{\dagger }(\mathbf{r})\delta \psi ^{\dagger }(\mathbf{r}^{\prime
})\right\rangle \left[\phi (\mathbf{r}^{\prime })\phi (\mathbf{r})+\frac{1}{2}%
\left\langle \delta \psi (\mathbf{r}^{\prime })\delta \psi (\mathbf{r}%
)\right\rangle \right]\bigg\},  \notag\\
E_{3B} &=g_{3}\int d\mathbf{r}\bigg[\frac{1}{3!}\left\vert \phi (\mathbf{r}%
)\right\vert ^{6}+\frac{3}{2}\left\vert \phi (\mathbf{r})\right\vert
^{4}\left\langle \delta \psi ^{\dagger }(\mathbf{r})\delta \psi (\mathbf{r}%
)\right\rangle +\text{Re}\left\vert \phi (\mathbf{r})\right\vert ^{2}\phi
^{2}(\mathbf{r})\left\langle \delta \psi ^{\dagger 2}(\mathbf{r}%
)\right\rangle\nonumber\\
&\quad +3\text{Re}\phi ^{2}(\mathbf{r})\left\langle \delta \psi
^{\dagger 2}(\mathbf{r})\right\rangle \left\langle \delta \psi ^{\dagger }(%
\mathbf{r})\delta \psi (\mathbf{r})\right\rangle+3\left\vert \phi (\mathbf{r})\right\vert ^{2}\left\langle \delta \psi
^{\dagger }(\mathbf{r})\delta \psi (\mathbf{r})\right\rangle ^{2}+\frac{3}{2}%
\left\vert \left\langle \delta \psi ^{2}(\mathbf{r})\right\rangle
\right\vert ^{2}\left\vert \phi (\mathbf{r})\right\vert ^{2}  \notag \\
&\quad+\left\langle
\delta \psi ^{\dagger }(\mathbf{r})\delta \psi (\mathbf{r})\right\rangle
^{3}+\frac{3}{2}\left\vert \left\langle \delta \psi ^{2}(\mathbf{r}%
)\right\rangle \right\vert ^{2}\left\langle \delta \psi ^{\dagger }(\mathbf{r%
})\delta \psi (\mathbf{r})\right\rangle \bigg]
\end{align}
being the kinetic, the two-body interaction, and the three-body interaction energies, respectively. The linear driving term is
\begin{align}
\eta({\mathbf r}) &=h_{0}\phi (\mathbf{r})\nonumber\\
&\quad+\int d\mathbf{r}^{\prime }U(\mathbf{r}-%
\mathbf{r}^{\prime })\bigg\{\Big[\left\vert \phi (\mathbf{r}^{\prime })\right\vert
^{2}+\left\langle \delta \psi ^{\dagger }(\mathbf{r}^{\prime })\delta \psi (%
\mathbf{r}^{\prime })\right\rangle \Big]\phi (\mathbf{r})\nonumber\\
&\qquad\qquad\qquad\qquad\quad+\left\langle \delta
\psi ^{\dagger }(\mathbf{r}^{\prime })\delta \psi (\mathbf{r})\right\rangle
\phi (\mathbf{r}^{\prime })+\left\langle \delta \psi (\mathbf{r})\delta \psi
(\mathbf{r}^{\prime })\right\rangle \phi ^{\ast }(\mathbf{r}^{\prime })\bigg\}
\notag \\
&\quad+g_{3}\bigg\{\frac{1}{2}\bigg[\left\vert \phi (\mathbf{r})\right\vert^{4}+\frac{1}{2}\left\langle \delta \psi ^{\dagger 2}(\mathbf{r})\right\rangle \phi ^{2}(\mathbf{r})+\frac{3}{2}\left\vert \left\langle
\delta \psi ^{2}(\mathbf{r})\right\rangle \right\vert ^{2}\nonumber\\
&\qquad\quad\;+3\left\langle \delta \psi ^{\dagger }(\mathbf{r})\delta \psi (\mathbf{r})\right\rangle \left(\left\vert \phi (\mathbf{r})\right\vert ^{2}+\left\langle
\delta \psi ^{\dagger }(\mathbf{r})\delta \psi (\mathbf{r})\right\rangle
\right)\bigg]\phi (\mathbf{r})\nonumber\\
&\qquad\quad\;  +3\left[\frac{1}{2}\left\vert \phi (\mathbf{r})\right\vert ^{2}+\left\langle
\delta \psi ^{\dagger }(\mathbf{r})\delta \psi (\mathbf{r})\right\rangle
\right]\left\langle \delta \psi ^{2}(\mathbf{r})\right\rangle \phi ^{\ast }(%
\mathbf{r})\bigg\}.
\end{align}
The mean-field Hamiltonian $\mathcal{H=}\left(
\begin{array}{cc}
\mathcal{E} & \Delta \\
\Delta ^{\dagger } & \mathcal{E}^{\ast }%
\end{array}%
\right) $ is determined by%
\begin{eqnarray}
\mathcal{E} &=&h_{0}+\delta (\mathbf{r}-\mathbf{r}^{\prime })\int d\mathbf{r}%
_{1}U(\mathbf{r}-\mathbf{r}_{1})[\left\vert \phi (\mathbf{r}_{1})\right\vert
^{2}+\left\langle \delta \psi ^{\dagger }(\mathbf{r}_{1})\delta \psi (%
\mathbf{r}_{1})\right\rangle ]+U(\mathbf{r}-\mathbf{r}^{\prime })[\phi
^{\ast }(\mathbf{r}^{\prime })\phi (\mathbf{r})+\left\langle \delta \psi
^{\dagger }(\mathbf{r}^{\prime })\delta \psi (\mathbf{r})\right\rangle ] \\
&&+\frac{3}{2}g_{3}\delta (\mathbf{r}-\mathbf{r}^{\prime })[\left\vert \phi (%
\mathbf{r})\right\vert ^{4}+4\left\vert \phi (\mathbf{r})\right\vert
^{2}\left\langle \delta \psi ^{\dagger }(\mathbf{r})\delta \psi (\mathbf{r}%
)\right\rangle +2\text{Re}\left\langle \delta \psi ^{\dagger 2}(\mathbf{r}%
)\right\rangle \phi ^{2}(\mathbf{r})+2\left\langle \delta \psi ^{\dagger }(%
\mathbf{r})\delta \psi (\mathbf{r})\right\rangle ^{2}+\left\vert
\left\langle \delta \psi ^{2}(\mathbf{r})\right\rangle \right\vert ^{2}]
\notag
\end{eqnarray}%
and%
\begin{eqnarray}
\Delta &=&U(\mathbf{r}-\mathbf{r}^{\prime })[\phi (\mathbf{r}^{\prime })\phi
(\mathbf{r})+\left\langle \delta \psi (\mathbf{r}^{\prime })\delta \psi (%
\mathbf{r})\right\rangle ] \\
&&+g_{3}\delta (\mathbf{r}-\mathbf{r}^{\prime })[\left\vert \phi (\mathbf{r}%
)\right\vert ^{2}\phi ^{2}(\mathbf{r})+3\phi ^{2}(\mathbf{r})\left\langle
\delta \psi ^{\dagger }(\mathbf{r})\delta \psi (\mathbf{r})\right\rangle
+3\left\vert \phi (\mathbf{r})\right\vert ^{2}\left\langle \delta \psi ^{2}(%
\mathbf{r})\right\rangle +3\left\langle \delta \psi ^{\dagger }(\mathbf{r}%
)\delta \psi (\mathbf{r})\right\rangle \left\langle \delta \psi ^{2}(\mathbf{%
r})\right\rangle ].  \notag
\end{eqnarray}

\section{Solving EOM by Hankel transformations}

In this section, we solve EOM (\ref{Imv}) projected in the angular momentum
basis by Hankel transformations. In the basis $e^{im\varphi }/\sqrt{2\pi }$,
the solution of Eq. (\ref{Imv}) has the form $\phi (\mathbf{r})=\phi
_{0}(\rho ,z)/\sqrt{2\pi }$ and%
\begin{eqnarray}
G(\mathbf{r,r}^{\prime }) &=&\frac{1}{2\pi }\sum_{m}e^{im(\varphi -\varphi
^{\prime })}G_{m}(\rho ,z\mathbf{;}\rho ^{\prime },z^{\prime }),  \notag \\
F(\mathbf{r,r}^{\prime }) &=&\frac{1}{2\pi }\sum_{m}e^{im(\varphi -\varphi
^{\prime })}F_{m}(\rho ,z\mathbf{;}\rho ^{\prime },z^{\prime }).
\end{eqnarray}%
The EOM for the wave function of coherent condensate reads%
\begin{equation}
\partial _{\tau }\phi _{0}(\rho ,z)=-\eta _{0}(\rho ,z)-2\int \rho ^{\prime
}d\rho ^{\prime }dz^{\prime }[G_{0}(\rho ,z;\rho ^{\prime },z^{\prime })\eta
_{0}(\rho ^{\prime },z^{\prime })+F_{0}(\rho ,z;\rho ^{\prime },z^{\prime
})\eta _{0}^{\ast }(\rho ^{\prime },z^{\prime })],
\end{equation}%
where%
\begin{eqnarray}
\eta _{0} &=&h_{0}^{m=0}\phi _{0}(\rho ,z)+\int \rho ^{\prime }d\rho
^{\prime }dz^{\prime }U_{0}(\rho ,z\mathbf{;}\rho ^{\prime },z^{\prime
})[\left\vert \phi _{0}(\rho ^{\prime },z^{\prime })\right\vert
^{2}+\sum_{m}G_{m}(\rho ^{\prime },z^{\prime }\mathbf{;}\rho ^{\prime
},z^{\prime })]\phi _{0}(\rho ,z)  \notag \\
&&+\sum_{m}\int \rho ^{\prime }d\rho ^{\prime }dz^{\prime }U_{m}(\rho ,z%
\mathbf{;}\rho ^{\prime },z^{\prime })[G_{m}(\rho ,z\mathbf{;}\rho ^{\prime
},z^{\prime })\phi _{0}(\rho ^{\prime },z^{\prime })+F_{m}(\rho ,z\mathbf{;}%
\rho ^{\prime },z^{\prime })\phi _{0}^{\ast }(\rho ^{\prime },z^{\prime
})]+\eta _{3}
\end{eqnarray}%
is determined by the contribution%
\begin{eqnarray}
\eta _{3} &=&\frac{g_{3}}{(2\pi )^{2}}\{\frac{1}{2}\left\vert \phi _{0}(\rho
,z)\right\vert ^{4}\phi _{0}(\rho ,z)+\frac{1}{2}\sum_{m}F_{m}^{\ast }(\rho
,z\mathbf{;}\rho ,z)\phi _{0}^{3}(\rho ,z)+\frac{3}{2}\left\vert
\sum_{m}F_{m}(\rho ,z\mathbf{;}\rho ,z)\right\vert ^{2}\phi (\rho ,z)  \notag
\\
&&+3\sum_{m}G_{m}(\rho ,z\mathbf{;}\rho ,z)[\left\vert \phi _{0}(\rho
,z)\right\vert ^{2}+\sum_{m^{\prime }}G_{m^{\prime }}(\rho ,z\mathbf{;}\rho
,z)]\phi _{0}(\rho ,z)  \notag \\
&&+3[\frac{1}{2}\left\vert \phi _{0}(\rho ,z)\right\vert
^{2}+\sum_{m}G_{m}(\rho ,z\mathbf{;}\rho ,z)]\sum_{m^{\prime }}F_{m^{\prime
}}(\rho ,z\mathbf{;}\rho ,z)\phi _{0}^{\ast }(\rho ,z)\}
\end{eqnarray}%
from the three-body interaction. The single-particle part%
\begin{equation}
h_{0}^{m}=-\frac{1}{2M}(\partial _{\rho }^{2}+\frac{1}{\rho }\partial _{\rho
}-\frac{m^{2}}{\rho ^{2}})-\frac{1}{2M}\partial _{z}^{2}+\frac{1}{2}M\omega
_{0}^{2}(\rho ^{2}+\lambda ^{2}z^{2})-\mu
\end{equation}%
and the interaction%
\begin{equation}
U_{m}(\rho ,z\mathbf{;}\rho ^{\prime },z^{\prime })=\int \frac{d\varphi
^{\prime }}{2\pi }U(\mathbf{r}-\mathbf{r}^{\prime })e^{-im(\varphi -\varphi
^{\prime })}
\end{equation}%
in the channel $m$ will be treated carefully by the Hankel transformation,
which can regularize the singularity in the dipolar interaction.

The EOM for the correlation functions are%
\begin{eqnarray}
\partial _{\tau }G_{m}(\rho ,z\mathbf{;}\rho ^{\prime },z^{\prime }) &=&-[(%
\mathcal{E}G)_{m}+(\Delta F^{\dagger })_{m}]-[(\mathcal{E}G)_{m}+(\Delta
F^{\dagger })_{m}]^{\dagger }  \notag \\
&&-2\int \rho _{1}d\rho _{1}dz_{1}\{G_{m}(\rho ,z\mathbf{;}\rho _{1},z_{1})[(%
\mathcal{E}G)_{m}+(\Delta F^{\dagger })_{m}](\rho _{1},z_{1};\rho ^{\prime
},z^{\prime })  \notag \\
&&+F_{m}(\rho ,z\mathbf{;}\rho _{1},z_{1})[(\Delta G^{T})_{-m}^{\ast }+(%
\mathcal{E}F)_{-m}^{\ast }](\rho _{1},z_{1};\rho ^{\prime },z^{\prime })\},
\end{eqnarray}%
and%
\begin{eqnarray}
\partial _{\tau }F_{m}(\rho ,z;\rho ^{\prime },z^{\prime }) &=&-\Delta _{%
\mathrm{3}}(\rho ,z)\frac{1}{\rho }\delta (\rho -\rho ^{\prime })\delta
(z-z^{\prime })-\Delta _{\mathrm{Bogo}}^{m}(\rho ,z;\rho ^{\prime
},z^{\prime })  \notag \\
&&-[(\mathcal{E}F)_{m}+(\Delta G^{T})_{m}]-[(\Delta G^{T})_{-m}+(\mathcal{E}%
F)_{-m}]^{T}  \notag \\
&&-2\int \rho _{1}d\rho _{1}dz_{1}\{G_{m}(\rho ,z\mathbf{;}\rho _{1},z_{1})[(%
\mathcal{E}F)_{m}+(\Delta G^{T})_{m}](\rho _{1},z_{1};\rho ^{\prime
},z^{\prime })  \notag \\
&&+F_{m}(\rho ,z\mathbf{;}\rho _{1},z_{1})[(\Delta F^{\dagger })_{-m}^{\ast
}+(\mathcal{E}G)_{-m}^{\ast }](\rho _{1},z_{1};\rho ^{\prime },z^{\prime
})\}.
\end{eqnarray}%
Here, we use the short-hand notations%
\begin{eqnarray}
(\mathcal{E}G)_{m}(\rho ,z;\rho ^{\prime },z^{\prime }) &=&[h_{0}^{m}+%
\mathcal{E}_{\mathrm{Hartree}}(\rho ,z)+\mathcal{E}_{\mathrm{3}}(\rho
,z)]G_{m}(\rho ,z\mathbf{;}\rho ^{\prime },z^{\prime })  \notag \\
&&+\int \rho _{1}d\rho _{1}dz_{1}\mathcal{E}_{\mathrm{Fock}}^{m}(\rho
,z;\rho _{1},z_{1})G_{m}(\rho _{1},z_{1}\mathbf{;}\rho ^{\prime },z^{\prime
}),  \notag \\
(\Delta F^{\dagger })_{m}(\rho ,z;\rho ^{\prime },z^{\prime }) &=&\Delta _{%
\mathrm{3}}(\rho ,z)F_{m}^{\dagger }(\rho ,z\mathbf{;}\rho ^{\prime
},z^{\prime })+\int \rho _{1}d\rho _{1}dz_{1}\Delta _{\mathrm{Bogo}%
}^{m}(\rho ,z;\rho _{1},z_{1})F_{m}^{\dagger }(\rho _{1},z_{1}\mathbf{;}\rho
^{\prime },z^{\prime }),
\end{eqnarray}%
and%
\begin{eqnarray}
(\mathcal{E}F)_{m}(\rho ,z;\rho ^{\prime },z^{\prime }) &=&[h_{0}^{m}+%
\mathcal{E}_{\mathrm{Hartree}}(\rho ,z)+\mathcal{E}_{\mathrm{3}}(\rho
,z)]F_{m}(\rho ,z\mathbf{;}\rho ^{\prime },z^{\prime })  \notag \\
&&+\int \rho _{1}d\rho _{1}dz_{1}\mathcal{E}_{\mathrm{Fock}}^{m}(\rho
,z;\rho _{1},z_{1})F_{m}(\rho _{1},z_{1}\mathbf{;}\rho ^{\prime },z^{\prime
}),  \notag \\
(\Delta G^{T})_{m}(\rho ,z;\rho ^{\prime },z^{\prime }) &=&\Delta _{\mathrm{3%
}}(\rho ,z)G_{-m}^{\ast }(\rho ,z\mathbf{;}\rho ^{\prime },z^{\prime })+\int
\rho _{1}d\rho _{1}dz_{1}\Delta _{\mathrm{Bogo}}^{m}(\rho ,z;\rho
_{1},z_{1})G_{-m}^{\ast }(\rho _{1},z_{1}\mathbf{;}\rho ^{\prime },z^{\prime
}),
\end{eqnarray}%
which are determined by the Hartree-Fock-Bogoliubov terms%
\begin{eqnarray}
\mathcal{E}_{\mathrm{Hartree}}(\rho ,z) &=&\int \rho _{1}d\rho
_{1}dz_{1}U_{0}(\rho ,z\mathbf{;}\rho _{1},z_{1})[\left\vert \phi _{0}(\rho
_{1},z_{1})\right\vert ^{2}+\sum_{m^{\prime }}G_{m^{\prime }}(\rho _{1},z_{1}%
\mathbf{;}\rho _{1},z_{1})],  \notag \\
\mathcal{E}_{\mathrm{Fock}}^{m}(\rho ,z;\rho ^{\prime },z^{\prime })
&=&\sum_{m^{\prime }}U_{m-m^{\prime }}(\rho ,z\mathbf{;}\rho ^{\prime
},z^{\prime })[\phi _{0}^{\ast }(\rho ^{\prime },z^{\prime })\phi _{0}(\rho
,z)\delta _{m^{\prime }0}+G_{m^{\prime }}(\rho ,z\mathbf{;}\rho ^{\prime
},z^{\prime })],  \notag \\
\Delta _{\mathrm{Bogo}}^{m}(\rho ,z;\rho ^{\prime },z^{\prime })
&=&\sum_{m^{\prime }}U_{m-m^{\prime }}(\rho ,z\mathbf{;}\rho ^{\prime
},z^{\prime })[\phi _{0}(\rho ^{\prime },z^{\prime })\phi _{0}(\rho
,z)\delta _{m^{\prime }0}+F_{m^{\prime }}(\rho ,z\mathbf{;}\rho ^{\prime
},z^{\prime })],
\end{eqnarray}%
and the contribution%
\begin{eqnarray}
\mathcal{E}_{\mathrm{3}}(\rho ,z) &=&\frac{3}{2}\frac{g_{3}}{(2\pi )^{2}}%
\{\left\vert \phi _{0}(\rho ,z)\right\vert ^{4}+4\left\vert \phi _{0}(\rho
,z)\right\vert ^{2}\sum_{m^{\prime }}G_{m^{\prime }}(\rho ,z\mathbf{;}\rho
,z)  \notag \\
&&+2\text{Re}\phi _{0}^{\ast 2}(\rho ,z)\sum_{m^{\prime }}F_{m^{\prime
}}(\rho ,z,\rho ,z)+2[\sum_{m^{\prime }}G_{m^{\prime }}(\rho ,z\mathbf{;}%
\rho ,z)]^{2}+\left\vert \sum_{m^{\prime }}F_{m^{\prime }}(\rho ,z,\rho
,z)\right\vert ^{2}\},  \notag \\
\Delta _{\mathrm{3}}(\rho ,z) &=&\frac{g_{3}}{(2\pi )^{2}}[\left\vert \phi
_{0}(\rho ,z)\right\vert ^{2}\phi _{0}^{2}(\rho ,z)+3\phi _{0}^{2}(\rho
,z)\sum_{m^{\prime }}G_{m^{\prime }}(\rho ,z\mathbf{;}\rho ,z)  \notag \\
&&+3\left\vert \phi _{0}(\rho ,z)\right\vert ^{2}\sum_{m^{\prime
}}F_{m^{\prime }}(\rho ,z,\rho ,z)+3\sum_{m^{\prime }m_{1}^{\prime
}}G_{m^{\prime }}(\rho ,z\mathbf{;}\rho ,z)F_{m_{1}^{\prime }}(\rho ,z,\rho
,z)]
\end{eqnarray}%
from the three-body interactions.

\section{Coupled GP-like equations}

In this section, we derive the coupled GP-like equations for coherent and squeezed condensates by minimizing the ground state energy $E_{0}$. For the squeezed coherent state ansatz with the spatial wave functions $\chi =(\bar{\phi}_{\mathrm{c}},\bar{\phi}_{\mathrm{s}})^{T}=(\sqrt{N_{\mathrm{c}}}\phi _{\mathrm{c}},\sqrt{N_{\mathrm{s}}}\phi _{\mathrm{s}})^{T}/\sqrt{N}$, the correlation functions are approximated as
\begin{eqnarray}
\left\langle \delta \psi ^{\dagger }(\mathbf{r}^{\prime })\delta \psi (%
\mathbf{r})\right\rangle &=&N\bar{\phi}_{\mathrm{s}}(\mathbf{r})\bar{\phi}_{%
\mathrm{s}}^{\ast }(\mathbf{r}^{\prime }),  \notag \\
\left\langle \delta \psi (\mathbf{r}^{\prime })\delta \psi (\mathbf{r}%
)\right\rangle &=&N\sqrt{1+N_{\mathrm{s}}^{-1}}\bar{\phi}_{\mathrm{s}}(%
\mathbf{r})\bar{\phi}_{\mathrm{s}}(\mathbf{r}^{\prime }),
\end{eqnarray}%
and the energy density $e=E/N$ reads%
\begin{equation}
e=\int d\mathbf{r[}\bar{\phi}_{\mathrm{c}}^{\ast }(\mathbf{r})h_{0}\bar{\phi}%
_{\mathrm{c}}(\mathbf{r})+\bar{\phi}_{\mathrm{s}}^{\ast }(\mathbf{r})h_{0}%
\bar{\phi}_{\mathrm{s}}(\mathbf{r})]+e_{2B}+e_{3B},
\end{equation}%
where%
\begin{eqnarray}
e_{2B} &=&N\int d\mathbf{r}d\mathbf{r}^{\prime }U(\mathbf{r}-\mathbf{r}%
^{\prime })[\frac{1}{2}\left\vert \bar{\phi}_{\mathrm{c}}(\mathbf{r}%
)\right\vert ^{2}\left\vert \bar{\phi}_{\mathrm{c}}(\mathbf{r}^{\prime
})\right\vert ^{2}+\left\vert \bar{\phi}_{\mathrm{s}}(\mathbf{r})\right\vert
^{2}\left\vert \bar{\phi}_{\mathrm{c}}(\mathbf{r}^{\prime })\right\vert ^{2}
\notag \\
&&+\frac{1}{2}(3+N_{\mathrm{s}}^{-1})\left\vert \bar{\phi}_{\mathrm{s}}(%
\mathbf{r})\right\vert ^{2}\left\vert \bar{\phi}_{\mathrm{s}}(\mathbf{r}%
^{\prime })\right\vert ^{2}+\bar{\phi}_{\mathrm{s}}^{\ast }(\mathbf{r})\bar{%
\phi}_{\mathrm{s}}(\mathbf{r}^{\prime })\bar{\phi}_{\mathrm{c}}^{\ast }(%
\mathbf{r}^{\prime })\bar{\phi}_{\mathrm{c}}(\mathbf{r})  \notag \\
&&+\sqrt{1+N_{\mathrm{s}}^{-1}}\text{Re}\bar{\phi}_{\mathrm{s}}^{\ast }(%
\mathbf{r})\bar{\phi}_{\mathrm{s}}^{\ast }(\mathbf{r}^{\prime })\bar{\phi}_{%
\mathrm{c}}(\mathbf{r}^{\prime })\bar{\phi}_{\mathrm{c}}(\mathbf{r})],
\end{eqnarray}%
and%
\begin{eqnarray}
e_{3B} &=&g_{3}N^{2}\int d\mathbf{r}[\frac{1}{3!}\left\vert \bar{\phi}_{%
\mathrm{c}}\right\vert ^{6}+\frac{3}{2}(3+N_{\mathrm{s}}^{-1})\left\vert
\bar{\phi}_{\mathrm{c}}\right\vert ^{2}\left\vert \bar{\phi}_{\mathrm{s}%
}\right\vert ^{4}+\frac{1}{2}(5+3N_{\mathrm{s}}^{-1})\left\vert \bar{\phi}_{%
\mathrm{s}}\right\vert ^{6}  \notag \\
&&+\frac{3}{2}\left\vert \bar{\phi}_{\mathrm{c}}\right\vert ^{4}\left\vert
\bar{\phi}_{\mathrm{s}}\right\vert ^{2}+\sqrt{1+N_{\mathrm{s}}^{-1}}\text{Re}%
\left\vert \bar{\phi}_{\mathrm{c}}\right\vert ^{2}\bar{\phi}_{\mathrm{c}}^{2}%
\bar{\phi}_{\mathrm{s}}^{\ast 2}+3\sqrt{1+N_{\mathrm{s}}^{-1}}\text{Re}\bar{%
\phi}_{\mathrm{c}}^{2}\bar{\phi}_{\mathrm{s}}^{\ast 2}\left\vert \bar{\phi}_{%
\mathrm{s}}\right\vert ^{2}].
\end{eqnarray}

The profiles of $\bar{\phi}_{\mathrm{c,s}}$ is determined by the imaginary
time evolution $\partial _{\tau }\chi =-\delta e/\delta \chi $. In the
explicit form,%
\begin{eqnarray}
\partial _{\tau }\bar{\phi}_{\mathrm{c}}(\mathbf{r}) &=&-h_{0}\bar{\phi}_{%
\mathrm{c}}(\mathbf{r})-N\int d\mathbf{r}^{\prime }U(\mathbf{r}-\mathbf{r}%
^{\prime })\{[\left\vert \bar{\phi}_{\mathrm{c}}(\mathbf{r}^{\prime
})\right\vert ^{2}+\left\vert \bar{\phi}_{\mathrm{s}}(\mathbf{r}^{\prime
})\right\vert ^{2}]\bar{\phi}_{\mathrm{c}}(\mathbf{r})  \notag \\
&&+[\bar{\phi}_{\mathrm{s}}^{\ast }(\mathbf{r}^{\prime })\bar{\phi}_{\mathrm{%
c}}(\mathbf{r}^{\prime })+\sqrt{1+N_{\mathrm{s}}^{-1}}\bar{\phi}_{\mathrm{c}%
}^{\ast }(\mathbf{r}^{\prime })\bar{\phi}_{\mathrm{s}}(\mathbf{r}^{\prime })]%
\bar{\phi}_{\mathrm{s}}(\mathbf{r})\}  \notag \\
&&-\frac{1}{2}g_{3}N^{2}\{[\left\vert \bar{\phi}_{\mathrm{c}}\right\vert
^{4}+6\left\vert \bar{\phi}_{\mathrm{c}}\right\vert ^{2}\left\vert \bar{\phi}%
_{\mathrm{s}}\right\vert ^{2}+\sqrt{1+N_{\mathrm{s}}^{-1}}\bar{\phi}_{%
\mathrm{c}}^{2}\bar{\phi}_{\mathrm{s}}^{\ast 2}+3(3+N_{\mathrm{s}%
}^{-1})\left\vert \bar{\phi}_{\mathrm{s}}\right\vert ^{4}]\bar{\phi}_{%
\mathrm{c}}  \notag \\
&&+3\sqrt{1+N_{\mathrm{s}}^{-1}}(\left\vert \bar{\phi}_{\mathrm{c}%
}\right\vert ^{2}\bar{\phi}_{\mathrm{s}}^{2}+2\left\vert \bar{\phi}_{\mathrm{%
s}}\right\vert ^{2}\bar{\phi}_{\mathrm{s}}^{2})\bar{\phi}_{\mathrm{c}}^{\ast
}\},
\end{eqnarray}%
and%
\begin{eqnarray}
\partial _{\tau }\bar{\phi}_{\mathrm{s}}(\mathbf{r}) &=&-h_{0}\bar{\phi}_{%
\mathrm{s}}(\mathbf{r})-N\int d\mathbf{r}^{\prime }U(\mathbf{r}-\mathbf{r}%
^{\prime })\{[\left\vert \bar{\phi}_{\mathrm{c}}(\mathbf{r}^{\prime
})\right\vert ^{2}+(3+N_{\mathrm{s}}^{-1})\left\vert \bar{\phi}_{\mathrm{s}}(%
\mathbf{r}^{\prime })\right\vert ^{2}]\bar{\phi}_{\mathrm{s}}(\mathbf{r})
\notag \\
&&+[\bar{\phi}_{\mathrm{c}}^{\ast }(\mathbf{r}^{\prime })\bar{\phi}_{\mathrm{%
s}}(\mathbf{r}^{\prime })+\sqrt{1+N_{\mathrm{s}}^{-1}}\bar{\phi}_{\mathrm{s}%
}^{\ast }(\mathbf{r}^{\prime })\bar{\phi}_{\mathrm{c}}(\mathbf{r}^{\prime })]%
\bar{\phi}_{\mathrm{c}}(\mathbf{r})\}  \notag \\
&&-\frac{1}{2}g_{3}N^{2}\{[3\left\vert \bar{\phi}_{\mathrm{c}}\right\vert
^{4}+6(3+N_{\mathrm{s}}^{-1})\left\vert \bar{\phi}_{\mathrm{c}}\right\vert
^{2}\left\vert \bar{\phi}_{\mathrm{s}}\right\vert ^{2}+3\sqrt{1+N_{\mathrm{s}%
}^{-1}}\bar{\phi}_{\mathrm{c}}^{\ast 2}\bar{\phi}_{\mathrm{s}}^{2}+3(5+3N_{%
\mathrm{s}}^{-1})\left\vert \bar{\phi}_{\mathrm{s}}\right\vert ^{4}]\bar{\phi%
}_{\mathrm{s}}  \notag \\
&&+\sqrt{1+N_{\mathrm{s}}^{-1}}(2\left\vert \bar{\phi}_{\mathrm{c}%
}\right\vert ^{2}\bar{\phi}_{\mathrm{c}}^{2}+9\left\vert \bar{\phi}_{\mathrm{%
s}}\right\vert ^{2}\bar{\phi}_{\mathrm{c}}^{2})\bar{\phi}_{\mathrm{s}}^{\ast
}\}.
\end{eqnarray}

\end{document}